\documentstyle[twocolumn,epsf]{jpsj}

\def\runtitle{Diamagnetic Response of Normal-Superconducting Double Layers}
\def\runauthor{A. {\sc Sumiyama}, T. {\sc Endo}, Y. {\sc Nakagawa}, and Y. {\sc Oda}}

\title
{
Diamagnetic Response of Normal-Superconducting Double Layers
}

\author
{ 
Akihiko {\sc Sumiyama}, Tomoko {\sc Endo}, Y\={o}hei {\sc Nakagawa}, and Yasukage {\sc Oda}
}

\inst
{
Department of Material Science, Faculty of Science, Himeji
Institute of Technology, Ak\={o}-gun 678-1297\\
}

\recdate
{
\hspace{1 cm}
}

\abst
{
The diamagnetism of a normal metal (N: Cu or Au), which is induced by the proximity effect of a superconductor (S: Nb), has been investigated for N-S double layers, which are formed by a thin-film deposition process. Detailed studies of samples, which have different electronic mean-free path $\ell_{{\rm N}}$ in N, suggest that $\ell_{{\rm N}}$ should be controlled by the impurity concentration rather than the mechanical imperfections in the lattice in order to clarify the $\ell_{{\rm N}}$ dependence of the proximity effect. Both the screening distance $\rho$ in N and the parameter $\nu$ in $\rho \propto T^{-\nu}$ increase with an increase in $\ell_{{\rm N}}$. This result can be understood on the assumption that the normal metal changes its behavior from the "dirty" limit ($\xi_{{\rm N}}>\ell_{{\rm N}}$) to the "clean" limit ($\xi_{{\rm N}}<\ell_{{\rm N}}$), where $\xi_{{\rm N}}$ is the coherence length in N.}

\kword
{
proximity effect, diamagnetism, Meissner effect, double layers
}
\begin{document}
\sloppy
\maketitle
%introduction
\section{Introduction}
Superconductivity is induced in a normal metal (N) by the proximity effect when in contact with a superconductor (S). Diamagnetic response of normal metals has been studied for N-S double layer systems in earlier investigations,\cite{1, 2} and then for N-clad S wires,\cite{3, 4, 5} because a good electrical contact between S and N is obtained easily in the latter wires.

From a theoretical point of view, proximity-induced diamagnetism of a normal metal is characterized by the electronic mean-free path $\ell_{{\rm N}}$ in the normal metal. Diamagnetic response has been calculated both in the dirty limit ($\ell_{{\rm N}}\to 0$)\cite{2} and in the clean limit ($\ell_{{\rm N}}\to \infty $).\cite{6} An attempt to introduce a parameter which characterizes the quality of an N-S interface has been done in both limits also.\cite{7, 8} Recently, the diamagnetic response of N-S double layers with arbitrary $\ell_{{\rm N}}$ has been studied theoretically in the case of a perfect N-S interface,\cite{9} and experimental results on N-clad S wires have been interpreted by use of that theory.\cite{10}

In N-clad S wires, a good electrical contact between N and S is obtained by a wire-drawing process, in which the area of the interface increases in proportion to the square root of the wire length.\cite{3} Consequently, the quality of the interface is enhanced greatly in thin wires. The drawing process, however, inevitably causes mechanical imperfections in the lattice in N, so that an annealing process is needed to improve $\ell_{{\rm N}}$. Since annealing at high temperatures tend to degrade the interface due to the interdiffusion between S and N,\cite{11} it is difficult to control sample properties, such as  $\ell_{{\rm N}}$, the quality of the interface, and the sample geometry, independently.

In this paper, N-S double layers, revived by improving the preparation process of the N-S interface, have been investigated. Since samples are made by the deposition of a superconductor {\it after} a normal-metal sheet, which has desired $\ell_{{\rm N}}$, is prepared, the $\ell_{{\rm N}}$ dependence of the proximity effect, when the quality of the N-S interface of double layers is the same, can be clarified experimentally.  

%Experimental
\section{Experimental}
The samples which we have investigated are N-S double layers, where S is Nb, and N is Cu or Au. Commercial Cu sheets and commercial Au sheets, which are 50 $\mu$m in thickness and of purity 4N and 5N, were cut up into strips 1 mm wide and 10 mm long. A Cu sheet of purity 6N and 50 $\mu$m in thickness, which was prepared by rolling out a commercial Cu plate 1 mm in thickness, was used also. Main impurities and source of each sheet are as follows. Cu sheet(4N): Bi+Sn+Mn+Sb+As+Se+Te(total) $<$40 ppm, S $<$15 ppm (Nilaco). Cu sheet(5N): Ca 4 ppm, Zn 4 ppm, Fe 3 ppm (Aldrich Chemical). Cu sheet(6N): Si 0.05 ppm, Ag 0.04 ppm, Al 0.02 ppm (Nilaco). Au sheet(4N): Ag 49 ppm, Cu 2 ppm, Fe 1 ppm (Tanaka Kikinzoku). Au sheet(5N): Fe 1 ppm, Ag $<$1 ppm, Cu $<$1 ppm (Tanaka Kikinzoku).

Some of the strips were annealed in an Ar atmosphere at a temperature of 600 $^{\circ }$C for one hour to remove the effect of cold work. Hereafter, samples are called, for example, "6N-a" or "6N", where "6N" denotes the purity, and "-a" denotes the annealed normal-metal. The electronic mean-free path $\ell_{{\rm N}}$ of normal-metal strips is calculated with the equation based on the free electron Fermi gas model,\cite{111}
\begin{equation}
\ell_{{\rm N}}=\frac{mv_{{\rm F}}}{ne^{2}\rho_{{\rm N}}}R. R.,
\label{eq0}
\end{equation}
where $m$ is the mass of a free electron, $n$ is the electron density, $\rho_{{\rm N}}$ is the electrical resistivity at the room temperature, $v_{{\rm F}}$ is the Fermi velocity in a normal metal, and $R. R.$ is the resistance ratio of the normal metal between the room temperature and 4.2 K, which is determined by resistance measurements along strips.
%Fig. 1
\begin{figure}[h]
\centerline{\epsfysize=4cm\epsfbox{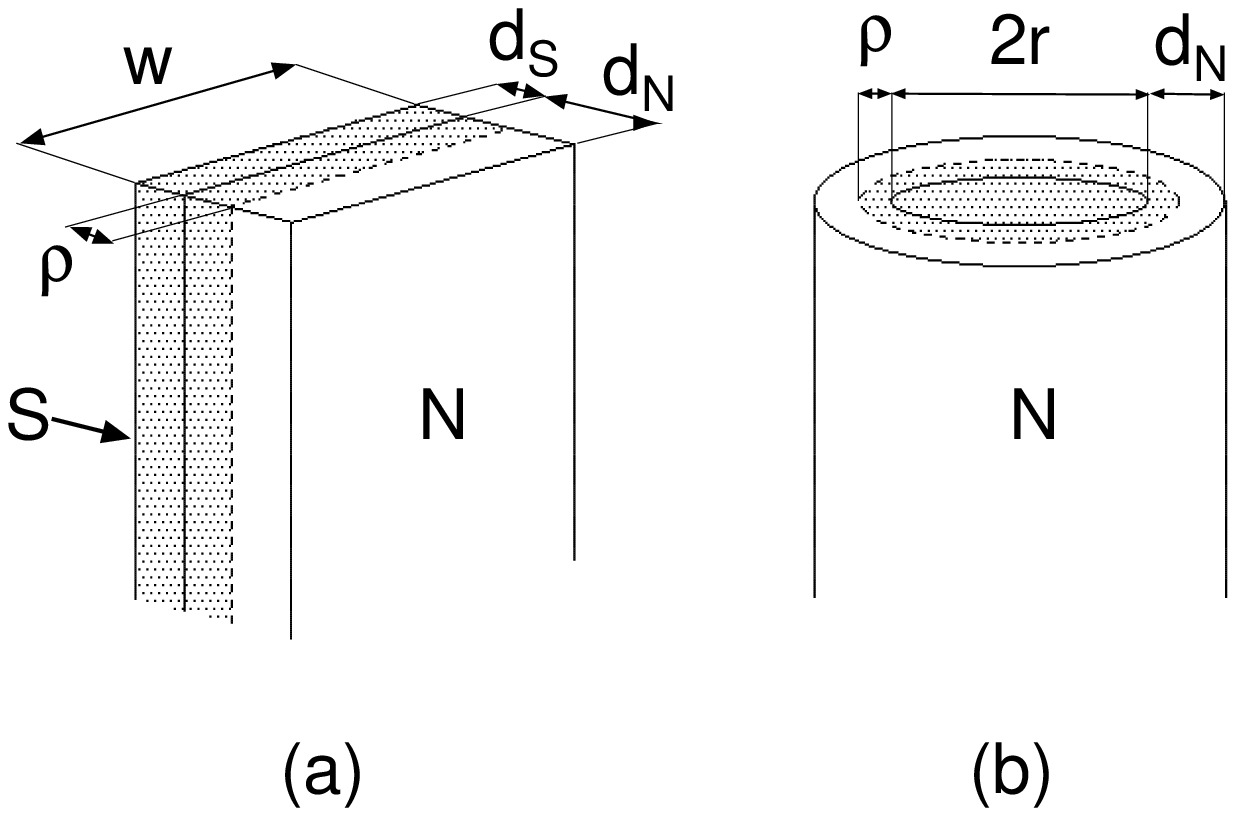}}
\caption{Schematic of (a) N-S double layers and (b) N-clad S wire. The shadowed area displays the Meissner effect.}
\label{fig:1}
\end{figure}

The surface of the normal-metal strips was rf sputter etched by Ar ion to a depth of about 2 $\mu$m, and then Nb (12 $\mu$m) was deposited by rf sputtering technique, as shown in Fig.~\ref{fig:1}(a). The temperatures of the strips were held at room temperature during the etching and deposition processes. In order to investigate the influence of the N-S geometry on the proximity effect, N-S double layers, of which width $w$ is half of the standard value (1 mm), or of which thickness $d_{{\rm S}}$ of Nb is half or twice as large as the standard value (12 $\mu$m), are made of Cu(6N-a) sheets. They are denoted as 6N-a1, 6N-a2, and 6N-a3, respectively.  
The thickness $d_{{\rm N}}$ of the normal-metal, $d_{{\rm S}}$, and $\ell_{{\rm N}}$ are listed in Table~\ref{table1}. 
%%%Table 1
\begin{table}[b]
\caption{Properties of samples.}
\label{table1}
\begin{tabular}{lccccc}
\hline
Sample&$r$ or $d_{{\rm S}}$&$d_{{\rm N}}$&$w$&$T_{{\rm anneal}}$&$\ell_{{\rm N}}$\\
&[$\mu$m]&[$\mu$m]&[mm]&[$^{\circ }$C]&[$\mu$m]\\
\hline
Cu(4N)&12&48&1.0&not&3.6\\
Cu(4N-a)&12&48&1.0&600&4.8\\
Cu(5N-a)&12&48&1.0&600&18\\
Cu(6N)&12&48&1.0&not&2.7\\
Cu(6N-a)&12&48&1.0&600&33\\
Cu(6N-a1)&12&48&0.5&600&33\\
Cu(6N-a2)&6&48&1.0&600&33\\
Cu(6N-a3)&24&48&1.0&600&33\\
Cu(clad)&125&50&-&600&3.5\\
Au(4N)&12&48&1.0&not&1.7\\
Au(4N-a)&12&48&1.0&600&11\\
Au(5N)&12&48&1.0&not&2.7\\
Au(5N-a)&12&48&1.0&600&15\\
Au(clad)&148&52&-&400&18\\
\hline
\end{tabular}
\end{table}

The N-S double layers were electrically insulated with varnish, and a bundle of about 30 strips was used for susceptibility measurements. The ac susceptibility was measured using an ac Hartshorn-type bridge by applying ac magnetic fields parallel to the strips. All measurements were performed at a frequency of 130 Hz in an ac field as low as 6 mOe. Varying the frequency between 30 and 270 Hz did not change our observations. The sample was mounted in a mutual inductance coil. It was connected to the mixing chamber of a dilution refrigerator and cooled down to 30 mK. The magnetic field in the sample region was reduced to less than 3 mOe by a $\mu$-metal shield. The temperature was measured by the carbon thermometers which were calibrated by a CMN thermometer.     
\begin{figure}[h]
\centerline{\epsfysize=4cm\epsfbox{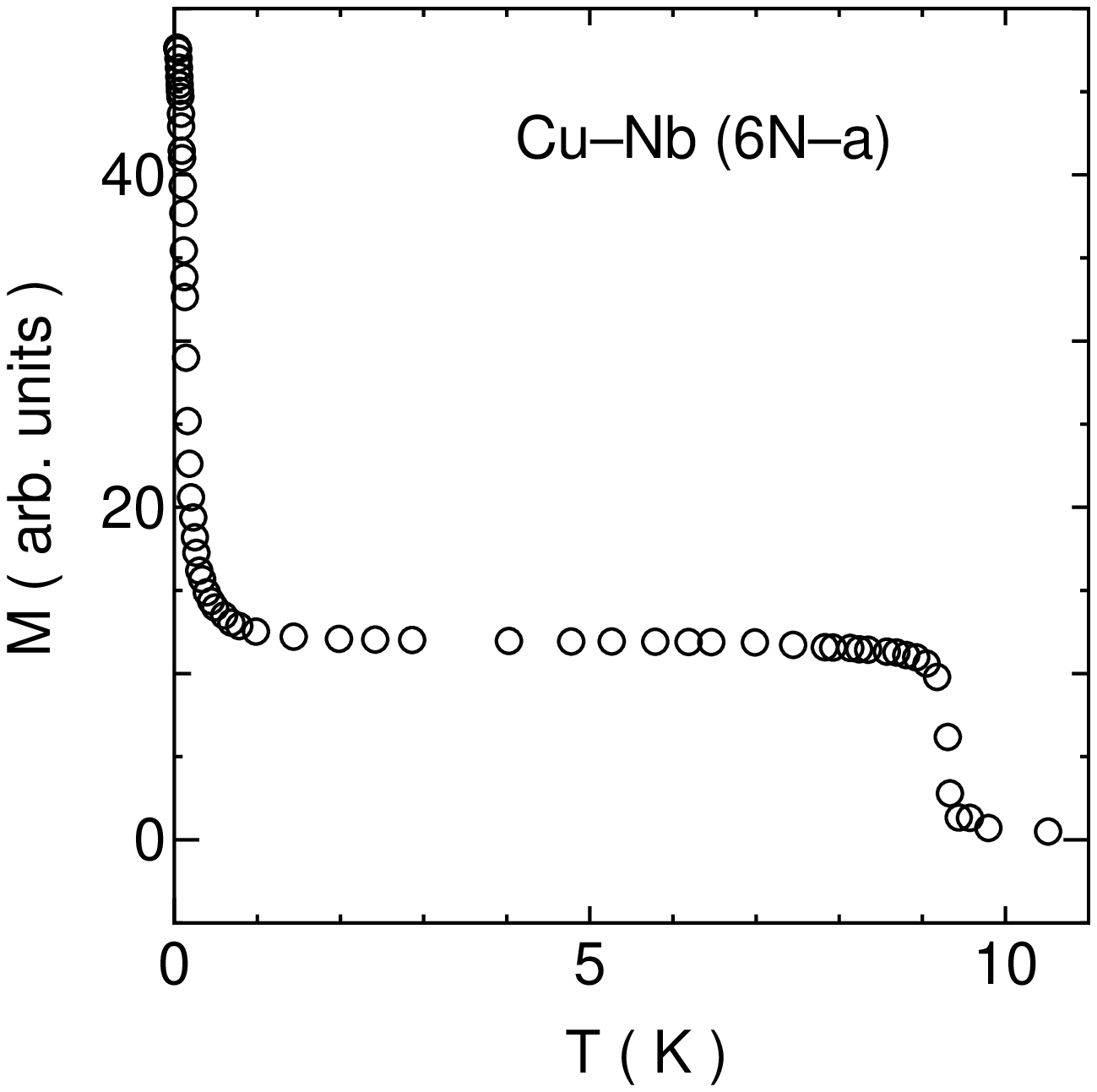}}
\caption{Typical temperature dependence of the mutual inductance $M$ measured in arbitrary units. As the temperature is decreased, the mutual inductance change due to the superconducting transition of Nb and then the proximity-induced diamagnetism of Cu is observed.}
\label{fig:12}
\end{figure}
%

%Results and Discussions
\section{Results and Discussion}
%Fig. 2
%Definition of rho
We show in Fig. \ref{fig:12} representative results for the temperature dependence of the mutual inductance $M$ when the sample 6N-a was mounted inside. When the temperature is decreased, a perfect diamagnetism of the Nb layer, followed by a nearly constant $M$, and then a further increase in  diamagnetism is observed. This implies that in some part of the normal-metal layer, which is in contact with Nb, the Meissner effect is induced by the proximity effect. The screening distance $\rho$ of the magnetic field in N is given by
\begin{equation}
\rho=d_{{\rm S}}\frac{\Delta M_{{\rm N}}}{\Delta M_{{\rm S}}},
\label{eq1}
\end{equation}
where $\Delta M_{{\rm S}}$ and $\Delta M_{{\rm N}}$ are the mutual inductance change due to the superconducting transition of Nb and the proximity-induced diamagnetism in the normal metal (Cu or Au), respectively. Considering that the superconducting transition of the Nb film in our samples is rather broad as compared with bulk Nb, and the temperature dependence of $M$ due to both Nb and the proximity effect is quite small near 4.2 K, we take $\Delta M_{{\rm S}}$=$M$(9.5 K)-$M$(4.2 K) and $\Delta M_{{\rm N}}$=$M$(4.2K)-$M$($T$). Since considerable care was taken to mount the specimens in the same way in the mutual inductance coil, the reproducibility of $\Delta M_{{\rm S}}$ and $\Delta M_{{\rm N}}$ measurements was better than 10\%. In addition, $\Delta M_{{\rm S}}$ and $\Delta M_{{\rm N}}$ tend to change in the same way by a small variation in the position of the specimens, so that the reproducibility of $\rho$, which is determined by the ratio $\Delta M_{{\rm N}}/\Delta M_{{\rm S}}$, was even better.

%Observation of the proximity effect 
%
%Fig. 3
\begin{figure}[h]
\centerline{\epsfysize=8cm\epsfbox{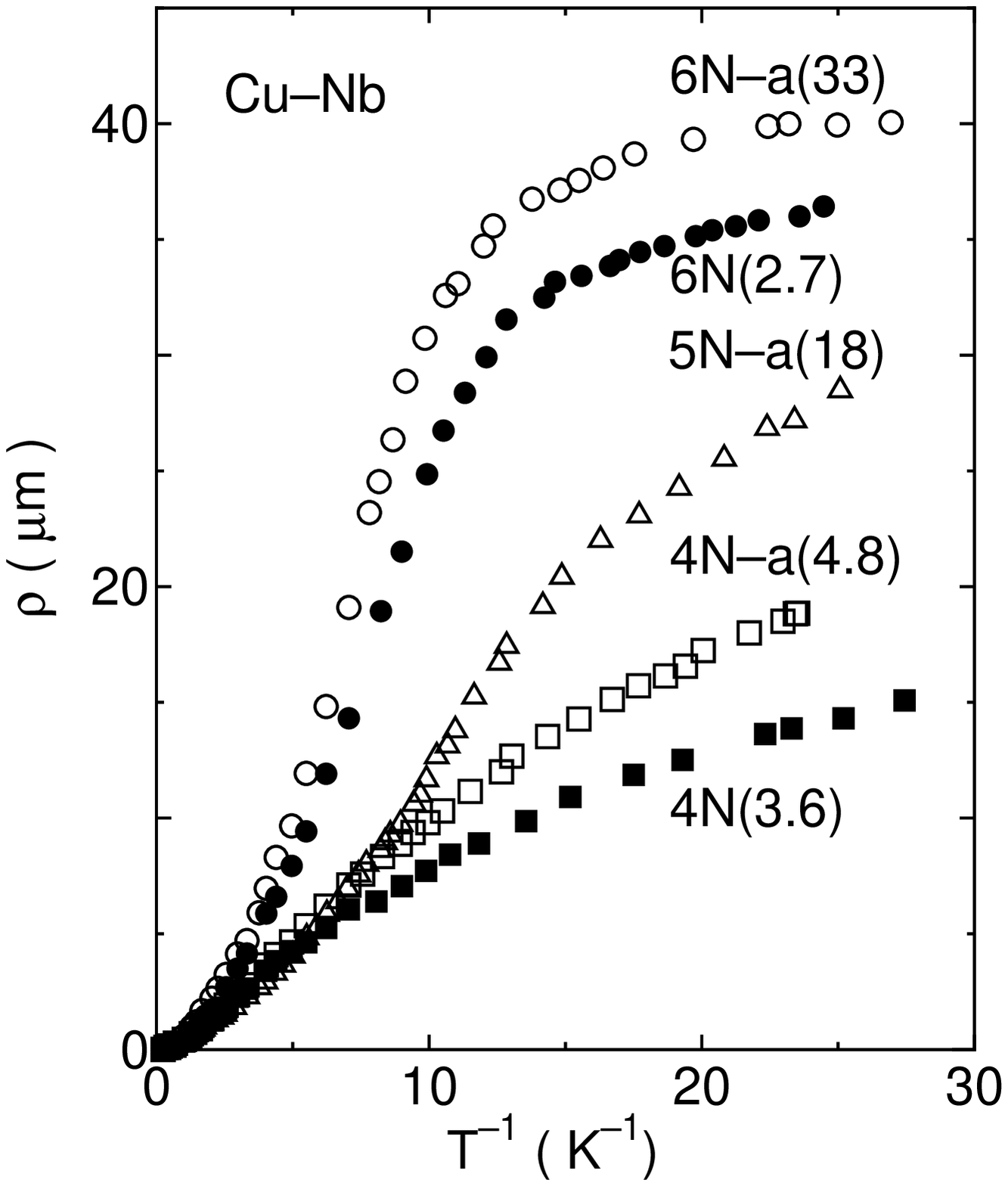}}
\caption{Temperature dependence of the screening distance $\rho$ for Cu-Nb double layers. The electronic mean free path $\ell_{{\rm N}}$ in Cu is referred to by the figures in parentheses.}
\label{fig:2}
\end{figure}
%
%%Fig. 4
\begin{figure}[h]
\centerline{\epsfysize=8cm\epsfbox{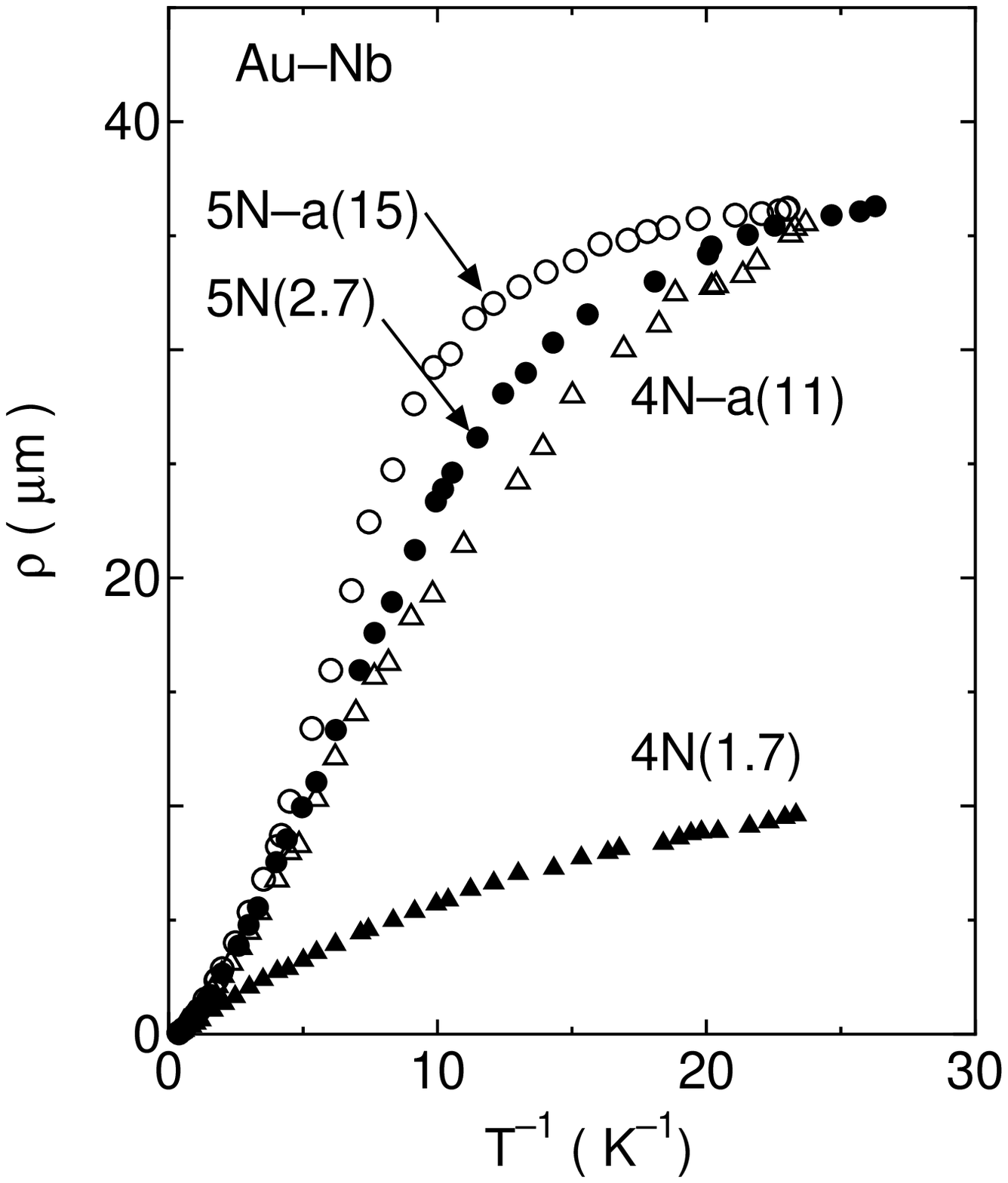}}
\caption{Temperature dependence of the screening distance $\rho$ for Au-Nb double layers. The electronic mean free path $\ell_{{\rm N}}$ in Au is referred to by the figures in parentheses.}
\label{fig:3}
\end{figure}
We show in Figs.~\ref{fig:2} and \ref{fig:3} the temperatures dependence of $\rho$ for Cu-Nb (Fig.~\ref{fig:2}) and Au-Nb (Fig.~\ref{fig:3}) double layers, which is plotted as $\rho$ vs. $T^{-1}$. Since the $\rho$ value depends on the amplitude $H_{{\rm ac}}$ of the ac magnetic field used for the measurement, the value in the limit of $H_{{\rm ac}}\to 0$ is plotted at every temperature. The proximity-induced superconductivity of Au, which was not observed in double layers such as Au-Sn in earlier works\cite{1, 12} and was found for the first time in Au-clad Nb wire,\cite{11} has been measured in the present double layers. This result gives some evidence that the interface between normal metals and Nb is sufficiently good.

As the purity of the normal metal is improved, $\rho$ becomes larger, and in the case of the same purity, $\rho$ is larger for the samples with annealed normal-metals. Samples which have relatively large $\rho$ exhibit saturation at low temperatures, indicating that the normal metal becomes fully superconducting. The screening distance $\rho_{0}$ at the low-temperature end, however, is between 37 and 40 $\mu$m, and is about 80\% of $d_{{\rm N}}$. Since $\rho$ of N-clad S wires exhibits saturation where $\rho$ is about $d_{{\rm N}}$, the deviation of $\rho_{0}$ from $d_{{\rm N}}$ in the case of N-S double layers may be ascribed to the sample geometry.

%Geometrical dependence of rho 
%%Fig. 5
\begin{figure}[h]
\centerline{\epsfysize=8cm\epsfbox{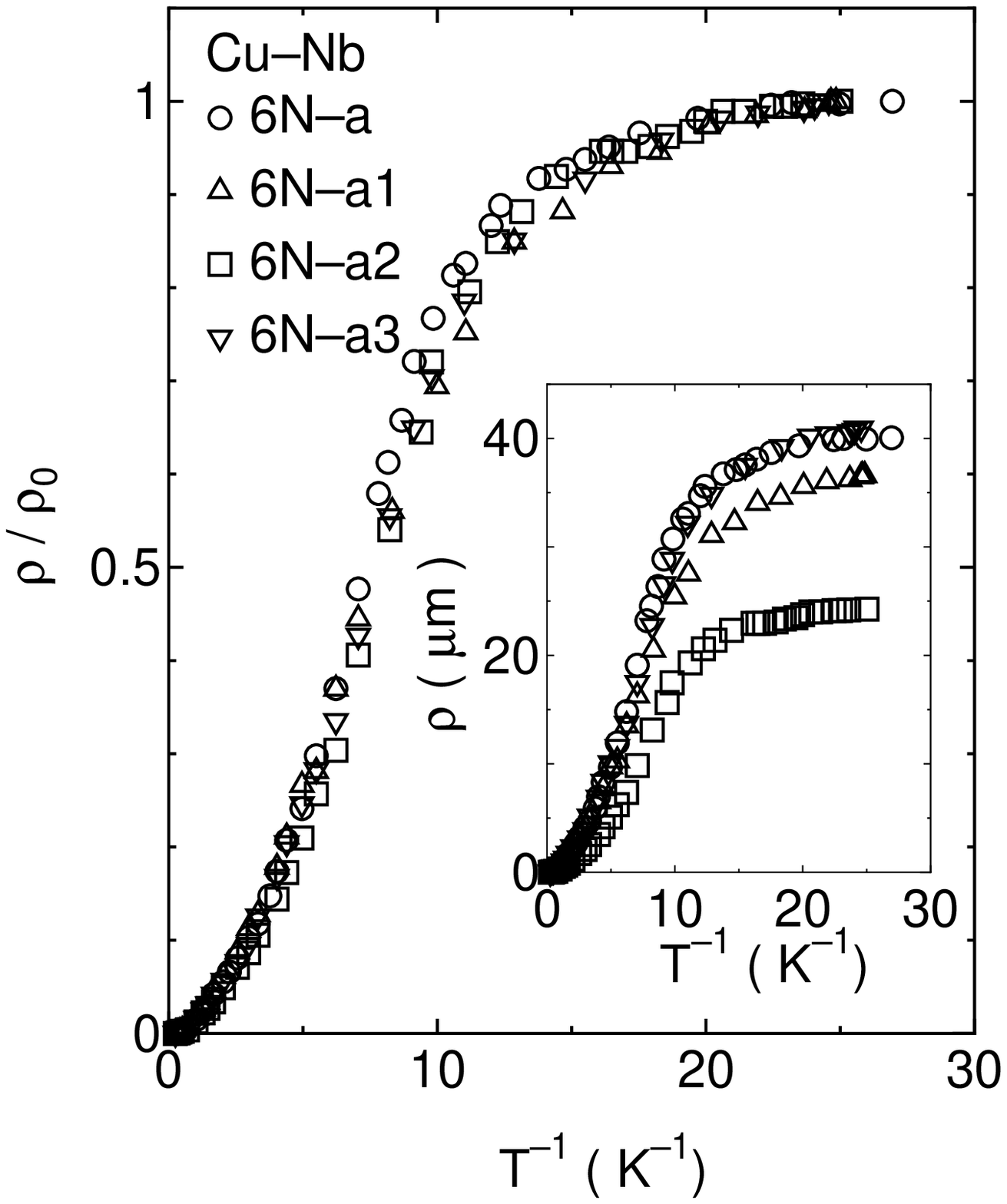}}
\caption{Temperature dependence of the ratio $\rho /\rho_{0}$ for four Cu-Nb samples made of the same Cu sheets (6N-a), where $\rho_{0}$ is $\rho$ at the low-temperature end. The inset shows the temperature dependence of $\rho$.}
\label{fig:4}
\end{figure}
In order to clarify the influence of the sample geometry on $\rho$, we have prepared Cu-Nb samples, of which width $w$ is half of the standard value (6N-a1), or of which $d_{{\rm S}}$ is half or twice as large as the standard thickness (6N-a2, 6N-a3), by use of the same Cu sheet (6N-a), as shown in Fig. \ref{fig:4}. The proximity effect of these samples is expected to be the same, since both $w$ and $d_{{\rm S}}$ is much larger than the coherence length or the penetration depth in S. The result that the ratio $\rho/\rho_{0}$ of the whole samples shows almost the same behavior suggests that the diamagnetic susceptibility of Cu is independent of $d_{{\rm S}}$ or $w$. Nevertheless, $\rho_{0}$ is between 37 and 40 $\mu$m, or even smaller in the case of $d_{{\rm S}}$=6 $\mu$m (6N-a2). It seems that $\rho$ tends to be  underestimated when eq.~(\ref{eq1}) is used; $\Delta M_{{\rm N}}/\Delta M_{{\rm S}}$ is smaller than the true $\rho/d_{{\rm S}}$.
%%%Table 2
\begin{table}[b]
\caption{Values of the ac susceptibilty change calculated from the superconducting transition signal of Nb strips, where $\Delta\chi_{{\rm Nb}}$  and $\Delta\chi_{{\rm Cu}}$ are the change due to the superconducting transition of Nb and the proximity-induced diamagnetism in Cu at the low-temperature end, respectively.}
\label{table2}
\begin{tabular}{lcccc}
\hline
Sample&$d_{{\rm S}}$&-4$\pi\Delta\chi_{{\rm Nb}}$&-4$\pi\Delta\chi_{{\rm Cu}}$&$\Delta\chi_{{\rm Cu}}/\Delta\chi_{{\rm Nb}}$\\
&[$\mu$m]&&&\\
\hline
Cu(6N-a)&12&0.94&0.79&0.83\\
Cu(6N-a2)&6&1.7&0.93&0.54\\
Cu(6N-a2)$^{a)}$&6&2.3&0.95&0.41\\
Cu(6N-a3)&24&1.0&0.84&0.84\\
\hline
\end{tabular}
a) The sample was inclined to the magnetic field at an angle of about 10$^{\circ}$ 
\end{table}

In order to relate $d_{{\rm S}}$ dependence of $\rho_{0}$  to the superconducting properties of Nb or Cu, we have measured the mutual inductance change at the superconducting transition of Nb strips 50 $\mu$m thick, 1 mm wide and 10 mm long as a calibration signal, and calculate the ac susceptibility change $\Delta\chi_{{\rm Nb}}$ from $\Delta M_{{\rm S}}$ and $\Delta\chi_{{\rm Cu}}$ from $\Delta M_{{\rm N}}$ at the low-temperature end, as listed in Table~\ref{table2}. It is obvious that $\rho_{0}$ calculated using eq.~(\ref{eq1}) is expressed as $\rho_{0}=d_{{\rm N}}\Delta\chi_{{\rm Cu}}/\Delta\chi_{{\rm Nb}}$ also. In the case of $d_{{\rm S}}$= 12 or 24 $\mu$m, Nb shows almost full diamagnetism (-$4\pi\Delta\chi_{{\rm Nb}}=1$) and $\rho_{0}$, which is about 80\% of $d_{{\rm N}}$, is ascribed to the small $\Delta\chi_{{\rm Cu}}$. As for $d_{{\rm S}}$= 6 $\mu$m, $\Delta\chi_{{\rm Nb}}$ is larger than that of full diamagnetism, which leads to even smaller $\rho_{0}$ value.

One possible explanation in the case of $d_{{\rm S}}$= 6 $\mu$m is related to the demagnetizing field; some of the strips happen to be placed along the direction which is slightly different from the magnetic field; $\Delta\chi_{{\rm Nb}}$ of those strips enhances due to the increase in demagnetizing field, while $\Delta\chi_{{\rm Cu}}$ does not increase so much as $\Delta\chi_{{\rm Nb}}$, because the proximity effect is reduced easily by the increase of magnetic field. The enhancement of $\Delta\chi_{{\rm Nb}}$ is probably remarkable, when the Nb layer is thin and $\Delta\chi_{{\rm Nb}}$ is rather small. This explanation, however, can not be applied to the present results quantitatively. We have inclined the sample 6N-a2 to the magnetic field at an angle of about 10$^{\circ}$ which is much larger than the angle variation expected when the sample is carefully mounted. Although $\Delta\chi_{{\rm Nb}}$ increases and $\rho_{0}$ decreases, the decrease of $\rho_{0}$ is smaller than the difference between $\rho_{0}$ and $d_{{\rm N}}$. The deviation of $\rho_{0}$ from $d_{{\rm N}}$ will be a subject of further investigations. Nevertheless, the above results suggest, at least, that $d_{{\rm S}}$ should be thicker than 12 $\mu$m so that $\rho$ may reflect only the proximity-induced superconductivity of Cu.  

%Temperature dependence
In addition to an increase in $\rho$ with eliminating impurities, a qualitative change in the temperature dependence is clearly seen in Figs.~\ref{fig:2} and \ref{fig:3}. If we approximately express the temperature dependence of $\rho$ as $\rho \propto T^{-\nu}$ at temperatures where $\rho$ is much smaller than $d_{{\rm N}}$, $\nu$ increases from $\nu<1$ to $1<\nu$ as the normal metal contains less impurities and approaches the "clean" limit; in other words, the second derivative $d^{2}\rho/(dT^{-1})^{2}$ changes from negative to positive.

In the dirty limit, $\rho$ is derived theoretically as follows,\cite{2}
\begin{equation}
\rho=\xi_{{\rm N}}[\ln (1/\kappa_{{\rm N}}(0))-0.116],
\label{eq2}
\end{equation}
where $\kappa_{{\rm N}}(0)$ is the local Ginzburg-Landau parameter in a normal metal at the N-S interface and $\xi_{{\rm N}}$ is the coherence length in N in the dirty limit ($\xi_{{\rm N}}\gg \ell_{{\rm N}}$), as given by
\begin{equation}
\xi_{{\rm N}}=(\hbar v_{{\rm F}}\ell_{{\rm N}}/6\pi k_{{\rm B}}
T)^{1/2},
\label{eq3}
\end{equation}
where $v_{{\rm F}}$ is the Fermi velocity in N. Since $\ln (1/\kappa_{{\rm N}}(0))$ depends only slightly on temperature, $\rho$ also shows the same temperature dependence $\rho \propto T^{-1/2}$ as $\xi_{{\rm N}}$. When $\ell_{{\rm N}}$ exceeds $\xi_{{\rm N}}$, the temperature dependence $\rho \propto T^{-1}$ may be expected using $\xi_{{\rm N}}$ in the clean limit ($\xi_{{\rm N}}\ll \ell_{{\rm N}}$), as given by
\begin{equation}
\xi_{{\rm N}}=\hbar v_{{\rm F}}/2\pi k_{{\rm B}}T.
\label{eq4}
\end{equation}

The result $\nu>1$ observed for Cu(6N-a), Cu(6N), Cu(5N-a), and Au(5N-a) indicates that eq.~(\ref{eq2}) can not be applied to these samples. This means that the local relation (London equation) between the diamagnetic current and the vector potential, on which eq.~(\ref{eq2}) is based, should be replaced by the non-local relation.\cite{6, 8} The diamagnetic response typical to the clean limit is seen in these samples. The decrease in $\nu$ from $\nu>1$ to $\nu<1$, which is observed for Cu(5N-a) with decreasing temperatures, reflects the fact that $\xi_{{\rm N}}\sim 1.9T^{-1}$($\mu$m$\cdot$K) in eq.~(\ref{eq4}) approaches $\ell_{{\rm N}}$ at $1/T\sim 10$. Previous studies of N-clad S wires have given various $\nu$ values: $\nu$=1/2,\cite{3} 1,\cite{3, 13} and $\nu>1$.\cite{4, 5} The present experiment on N-S double layers proves that $\nu$ in $\rho \propto T^{-\nu}$ increases through $\nu$=1 as the normal metal becomes cleaner.

%Mean free path dependence
The present results agree with the theory in ref. 9, on the assumption that $\ell_{{\rm N}}$ increases successively below $d_{{\rm N}}$ with a decrease of impurity concentration and a decrease of mechanical imperfections in the lattice by annealing. The $\ell_{{\rm N}}$ values determined experimentally, however, are not in order of $\rho$; Cu(6N) has the smallest $\ell_{{\rm N}}$ among Cu-Nb samples, and $\ell_{{\rm N}}$ of Au(5N) is smaller than $\ell_{{\rm N}}$ of Au(4N-a). This means that large $\rho$ is observed in spite of small $\ell_{{\rm N}}$ in non-annealed normal-metals which contain many imperfections in the lattice.

The present result may be interpreted several ways: (1) The distribution of mechanical imperfections in the lattice may not be uniform; there may be some regions which have high resistivity. Then, the resistance ratio $R. R. $ and the measured $\ell_{{\rm N}}$ become small, while the proximity effect occurs in the low-resistivity region. (2) Even if the distribution of the imperfections is uniform, the anisotropy of resistivity may exist, since the imperfections appear by the strong cold work along one direction, such as the rolling-out process (N-S double layers) or the wire-drawing process (N-clad S wires). In case that they are not removed by annealing, the $\ell_{{\rm N}}$ value measured for the current direction along the strip may not directly characterize the proximity effect caused by the penetration of Cooper pairs perpendicular to the N-S interface. So far, the anisotropy of resistivity has not been determined yet because of the small resistance for the current direction perpendicular to the strip surface. Furthermore, there is no microscopic information about the mechanical imperfections in the lattice. Still, when the complicated influence of the imperfections on the proximity effect is taken into account, the effect of cold work should be removed by annealing and $\ell_{{\rm N}}$ should be controlled by the impurity concentration alone in order to investigate the $\ell_{{\rm N}}$ dependence of $\rho$.  

%Au-clad Nb wires 
The difficulty to control $\ell_{{\rm N}}$ by imperfections in the lattice may also explain the discrepancy between the present result for Au-Nb double layers and the previous result for Au-clad Nb wires. In ref. 14, we prepared Au-clad Nb wires with $\ell_{{\rm N}}$= 0.11$d_{{\rm N}}$, 0.14$d_{{\rm N}}$, and 0.34$d_{{\rm N}}$ by annealing the same wires at different temperatures, and observed the same behavior of $\rho$. We have explained the result by use of eq.~(\ref{eq2}) with a modification replacing $\xi_{{\rm N}}$ in the dirty limit (eq.~(\ref{eq3})) with $\xi_{{\rm N}}$ in the clean limit (eq.~(\ref{eq4})); $\rho$ is independent of $\ell_{{\rm N}}$ like $\xi_{{\rm N}}$. The present result that $\rho$ increases when $\ell_{{\rm N}}$ is increased, however, contradicts this explanation. It is likely that annealing Au-clad Nb wires at different temperatures, which have changed the measured $\ell_{{\rm N}}$ values, have little effect on the mean-free path which characterizes the proximity effect. 

%The quality of the N-S interface
On the basis of the importance to control $\ell_{{\rm N}}$ only by impurities, we shall, for the moment, confine ourselves to the data on annealed normal-metals. The clean-limit behavior $\nu>1$ in $\rho \propto T^{-\nu}$ is observed at higher temperatures for Cu(6N-a), Cu(5N-a), and Au(5N-a), of which $\ell_{{\rm N}}\ge$ 15 $\mu$m, while $\nu\le 1$ for Cu(4N-a) and Au(4N-a), of which $\ell_{{\rm N}}\le$ 11 $\mu$m. This result suggests that the temperature dependence is determined mostly by $\ell_{{\rm N}}$. On the other hand, the $\rho$ values of Au(5N-a) and Au(4N-a) are larger than Cu(5N-a) in spite of smaller $\ell_{{\rm N}}$, suggesting that the quality of the N-S interface determines the $\rho$ values; the quality of the Cu-Nb interface is inferior to that of the Au-Nb interface.
%%Fig. 6
\begin{figure}[h]
\centerline{\epsfysize=8cm\epsfbox{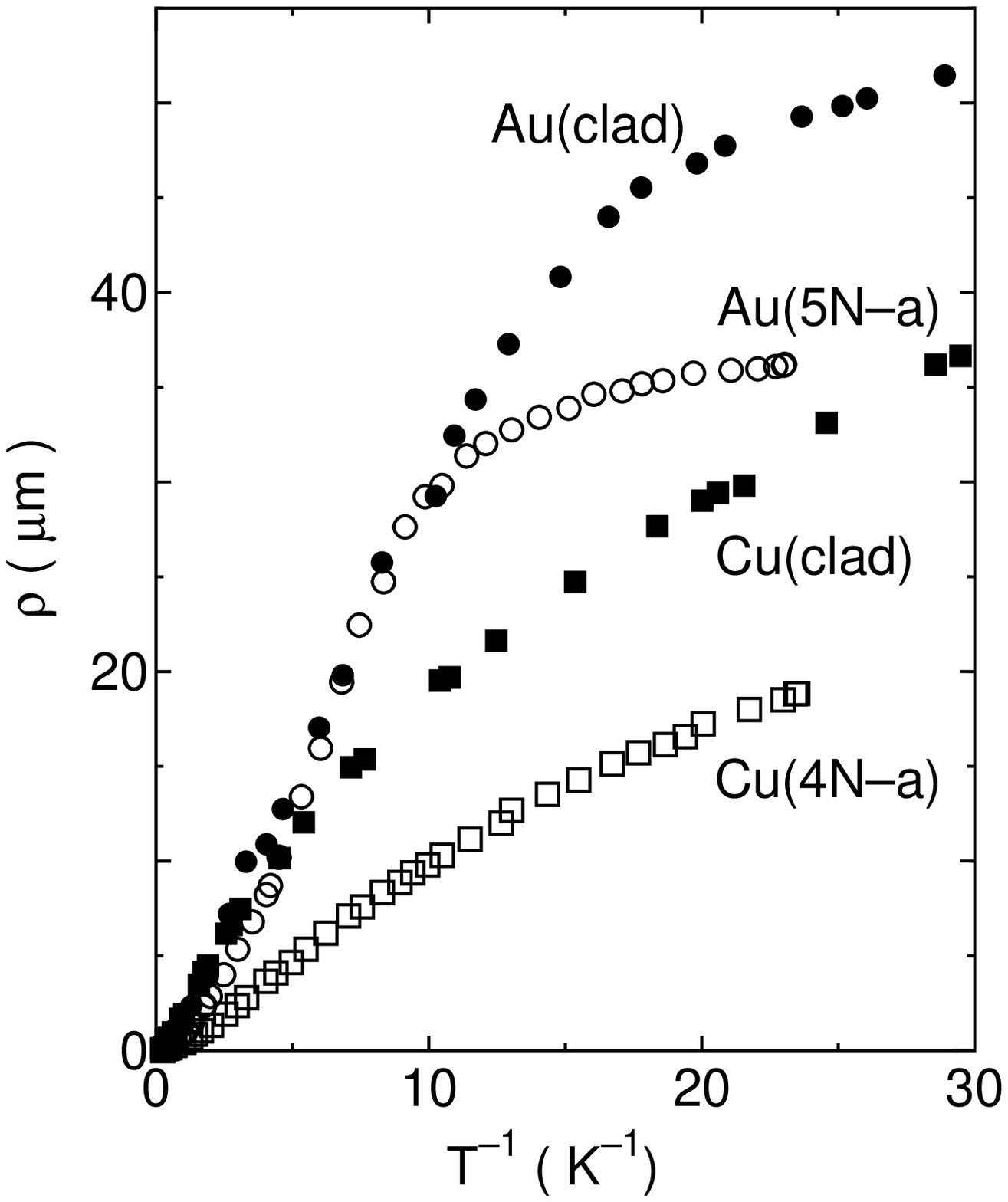}}
\caption{Comparison of $\rho$ between N-S double layers and N-clad S wires, which have almost the same $\ell_{{\rm N}}$. }
\label{fig:5}
\end{figure}

The difference in the quality of the N-S interface is obvious when we compare the present result with the data on N-clad S wires. Figure \ref{fig:5} shows the screening distance $\rho$ of Cu(4N-a), Au(5N-a), and two N-clad S wires: one is the Cu-clad Nb wire\cite{3} of which $\ell_{{\rm N}}$ is a little smaller than that of Cu(4N-a), and the other the Au-clad Nb wire\cite{13} which has a little larger $\ell_{{\rm N}}$ than Au(5N-a). Even if $\rho$ of N-S double layers is probably underestimated by a factor of 0.8, as mentioned above, it is obvious that $\rho$ is reduced in Cu-Nb double layers due to the lower quality of Cu-Nb interface as compared with Cu-clad Nb wires. On the other hand, the quality of Au-Nb interface in Au-Nb double layers is comparable to that in Au-clad Nb wires. Since Nb/Cu has a low mutual solubility\cite{14} and Nb/Au dissolves each other for temperatures above 500$^{\circ}$C\cite{15}, the degradation of the interface during the deposition of Nb is thought to occur more easily in Au-Nb double layers. The above result may suggest that some other properties, such as the quality of the etched surface of Cu or Au, are more important to obtain high-quality N-S interfaces.  

%Conclusion
\section{Conclusion}
In conclusion, we have prepared normal(N: Cu or Au)-superconducting(S: Nb) double layers, which have different mean-free path $\ell_{{\rm N}}$ in N, and yet have the N-S interface of the same quality, and investigated the proximity-induced diamagnetism of N. Systematic studies of the $\ell_{{\rm N}}$ dependence of the screening distance $\rho$ in N suggest the importance to control $\ell_{{\rm N}}$ by the impurity concentration rather than the mechanical imperfections in the lattice. Both $\rho$ and the parameter $\nu$ in $\rho \propto T^{-\nu}$ increase with an increase in $\ell_{{\rm N}}$. This result can be understood on the assumption that the normal metal changes its behavior from the "dirty" limit ($\xi_{{\rm N}}>\ell_{{\rm N}}$) to the "clean" limit ($\xi_{{\rm N}}<\ell_{{\rm N}}$). Although $\rho$ is sensitive to the quality of the N-S interface, it seems that $\nu$ is determined mainly by $\ell_{{\rm N}}$.

The present process to make the N-S proximity system is also useful, when S or N is a material which is difficult to use for the wire-drawing process. An attempt to measure the proximity effect of unconventional superconductors, such as heavy-fermion superconductors, is in progress.

%Acknowledgement
We would like to thank Prof. J. Hara, Dr. S. Higashitani, and Prof. K. Nagai for helpful discussions. 

\end{document}